\documentclass[12pt,twoside]{article}

\newcommand{\rd}{\mathrm{d}}
\newcommand{\rP}{{\mathrm{P}}}
\newcommand{\rS}{{\mathrm{SCRAM}}}
\newcommand{\rB}{{\mathrm{B}}}
\newcommand{\half}{{\textstyle{\frac12}}}

\input BoxedEPS
\SetRokickiEPSFSpecial  
\HideDisplacementBoxes

\newcommand{\numFig}[3]{\begin{figure}[ht]
 $$\BoxedEPSF{#3}$$
    \caption{#2}\label{fig:#1}
   \end{figure}}
\newcommand{\refFig}[1]{Figure~\ref{fig:#1}}

\begin{document}
\title{A Numerical Study of the Performance of a
Quantum Adiabatic Evolution Algorithm  for
Satisfiability}
\author{Edward Farhi, Jeffrey
Goldstone
\\ 
\small Center for Theoretical Physics \\[-1ex]  
\small  Massachusetts Institute of Technology \\[-1ex]    
\small  Cambridge, MA 02139\\[-1ex]   
\small  {\tt farhi@mit.edu} \qquad  {\tt goldstone@mitlns.mit.edu}\\[2ex] 
Sam Gutmann \\
\small Department of Mathematics \\[-1ex]   
\small  Northeastern University \\[-1ex]    
\small  Boston, MA 02115\\[-1ex]
\small {\tt sgutm@neu.edu}
}
\date{\footnotesize\sf MIT-CTP~\#3006} 
\maketitle
\pagestyle{myheadings}
\markboth{Edward Farhi, Jeffrey
Goldstone, and Sam Gutmann}{Performance of a 
Quantum Adiabatic Evolution Algorithm  for
Satisfiability}
\vspace*{-2.75pc}

\begin {abstract}\noindent
Quantum computation by adiabatic evolution, as
described in quant-ph/0001106, will solve 
satisfiability problems if the running time is long
enough.  In certain special cases (that are classically easy)
we know that the quantum algorithm requires a
running time that grows as a  polynomial in the
number of bits.  In this paper we present numerical
results on randomly generated instances of an NP-complete problem and of a
problem that can be solved classically in polynomial time. We simulate a
quantum computer (of up to 16 qubits) by integrating the Schr\"{o}dinger
equation on a conventional computer. For both problems considered, for the set
of instances studied, the required running time appears to grow slowly as a
function of the number of bits. 
\end{abstract}

\setcounter{equation}{0}
\section{\large Introduction}
\label{sec:1}

A quantum algorithm for the satisfiability problem was presented in [1].  This
algorithm is based on quantum adiabatic evolution.  If a state $|\psi(t)\rangle$
evolves according to the Schr\"{o}dinger equation with a slowly varying
Hamiltonian $H(t)$ and $|\psi(0)\rangle$ is the ground state of $H(0)$, then
$|\psi(t)\rangle$ will stay close to the instantaneous ground state of $H(t)$. 
The Hamiltonian $H(t)$ used in the algorithm is designed so that the ground
state of
$H(0)$ is easy to construct and the ground state of $H(T)$ encodes the solution
to the instance of satisfiability.  A crucial question is how large must the
running time, $T$, be to achieve an acceptable probability of success.  In this
paper we simulate an $n$-qubit continuous time quantum computer by
numerically integrating the Schr\"{o}dinger equation in a $2^n$-dimensional
Hilbert space.  We randomly generate difficult instances of an NP-complete
problem and study how large $T$ must be as a function of the number of bits
$n$.  For
$7\leq n\leq 16$, we find that the required $T$ grows modestly with $n$ since
the data is well fit by a quadratic in $n$.

\setcounter{equation}{0}
\section{\large A quantum algorithm based on 
adiabatic evolution}
\label{sec:2}

An $n$-bit instance of satisfiability is a Boolean formula
with M clauses
\begin{equation}
C_1 \wedge C_2  \wedge \cdots \wedge C_M 
\label{21}
\end{equation}
where each clause $C_a$ is True or False depending
on the values of some subset of the $n$ bits.  The task is to
discover if one (or more) of the $2^n$ possible
assignments of the values of the $n$ bits satisfies all of the
clauses, that is, makes formula~(\ref{21})
True. We consider two restricted versions of satisfiability, both of which are
restricted versions of a problem called ``exact cover".
In the first, EC3, each clause involves only three
bits.   The clause always has the same form.  The
clause is True if and only if one of the three bits
is a $1$ and the other two are $0$, so there are
three satisfying assignments out of the eight
possible values of the three bits. The second, EC2, has the restriction that
each clause involves only two bits.  In this case the clause is True if and
only if the two bits have the value 01 or 10, so there are two satisfying
assignments out of the four possible values for the two bits.
We pick these two
examples because EC2 is classically solvable in polynomial time 
whereas EC3 is NP-complete and no polynomial
time algorithm is known.  It is interesting to see how the quantum algorithm
treats these two classically very different problems.

To understand the quantum algorithm we first
recall the content of the adiabatic theorem.  We
are given a Hamiltonian $\tilde{H}(s)$ that
depends smoothly on the parameter $s$ where $s$
varies from $0$ to $1$.  Suppose that for each
value of $s$, $\tilde{H}(s)$ has a unique lowest
energy state, the ground state $|g;s\rangle$. That is,  
\begin{equation}
\tilde{H}(s)|g;s\rangle \,=\, E_0(s)|g;s\rangle
\label{22}
\end{equation}
where $E_0(s)$ is strictly less than any of the
other eigenvalues of $\tilde{H}(s)$.  Introduce a
time scale $T$ and define a time-dependent
Hamiltonian ${H}(t)$ by 
\begin{equation}
H(t) \,=\,\tilde{H}(t/T)
\label{23}
\end{equation}
where $t$ varies from $0$ to $T$. As $T$ gets
bigger, $H(t)$ becomes more slowly varying.  Let
$|\psi(t) \rangle$ obey the Schr\"{o}dinger equation
\begin{equation}
i \frac {\rd}{\rd t}|\psi(t)\rangle
\,=\,H(t)|\psi(t)\rangle
\label{24}
\end{equation}
with 
\begin{equation}
|\psi(0)\rangle \,=\,|g; s=0\rangle.
\label{25}
\end{equation}
That is, at $t=0$, $|\psi(0)\rangle$ is the ground
state of $H(0)$.  The adiabatic theorem tells us
that 
\begin{equation}
\lim_{T \to\infty} \bigl| \langle g;
s=1 | \psi(T)\rangle
\bigr|=1 \ .
\label{26}
\end{equation}
This means that for $T$ large enough,
$|\psi(T)\rangle$ is (up to a phase) very close to
the ground state of $H(T)$.  Eq.~(\ref{26}) only
holds if the gap between the ground state energy,
$E_0(s)$ of $\tilde{H}(s)$, and the next highest
energy, $E_1(s)$ of $\tilde{H}(s)$, is strictly
greater than zero, that is, $E_1(s)-E_0(s)>
0$ for $0\leq s\leq 1$. We will also discuss cases where at
$s=1$ there are multiple ground states. In this situation the
evolution ends very close to the ground state subspace. 

The idea behind the quantum algorithm is the
following.  Given an instance of the satisfiability
problem it is straightforward to construct a
Hamiltonian, $H_\rP$, whose ground state corresponds
to an assignment of the bits that violates the
least number of clauses.  Although it is easy to
construct $H_\rP$, finding its ground state may be
computationally challenging.  However we can
construct a Hamiltonian $H_\rB$ whose ground state we
explicitly know.  Now let
\begin{equation}
\tilde{H}(s)=(1-s)H_\rB +s H_\rP
\label{27}
\end{equation}
and accordingly
\begin{equation}
H(t)=(1-t/T)H_B+(t/T)H_\rP
\label{28}
\end{equation}
for some fixed $T$.  We start our quantum system in
the known ground state of $H_\rB$ and then evolve
according to Eq.~(\ref{24}) for time $T$. Suppose that the instance of
satisfiability that gave rise to $H_\rP$ has a unique satisfying
assignment. If $T$ is large enough, then
$|\psi(T)\rangle$ will be very close to the ground
state of $H_\rP$.  Measuring $|\psi(T)\rangle$ will,
with high probability, produce the satisfying
assignment. In general this algorithm will produce
an assignment that violates the minimum number of
clauses in Eq.~(\ref{21}).  Restricting to
instances with a unique satisfying assignment
appears to pick out  difficult instances
and simplifies the analysis of our algorithm.

More explicitly, given an $n$-bit instance of
satisfiability, we work in an $n$-qubit Hilbert
space that is spanned by the $2^n$ basis vectors
$|z_1\rangle |z_2\rangle\cdots |z_n\rangle$ where
$z_i=0,1$ and $|z_i\rangle$ is an eigenstate of the
$z$ component of the $i^{th}$ spin, 
\begin{equation}
\half (1-\sigma^{(i)}_z) |z_i\rangle = z_i|z_i\rangle
\quad \mbox{ where each}\quad
\sigma^{(i)}_z = \Biggl(\begin{array}{cc}1 & 0\\0 & -1\end{array}\Biggr)\, . 
\label{29}
\end{equation}
We also need the eigenstates of the $x$ component of
the $i^{th}$ spin,
$$
|x_i=0\rangle=\frac {1}{\sqrt{2}}(|z_i=0\rangle
+|z_i=1\rangle)
$$ 
and
\begin{equation}
|x_i=1\rangle=\frac {1}{\sqrt{2}}(|z_i=0\rangle
-|z_i=1\rangle)
\label{210}
\end{equation}
that obey
\begin{equation}
\half (1-\sigma_x^{(i)}) \left|x_i=x\right\rangle=x\left|x_i=x\right\rangle\quad
\mbox{ where each }\quad
\sigma_x^{(i)} = \Biggl(\begin{array}{cc}0 & 1\\1 & 0\end{array}\Biggr)\,.
\label{211}
\end{equation}

For concreteness imagine  that each clause in
formula (\ref{21}) involves $3$ bits.  Let
$i_C,j_C$ and $k_C$ be the $3$ bits associated with
clause $C$.  For each clause $C$ define an
``energy" function
\begin{eqnarray}
h_C(z_{i_C},z_{j_C},z_{k_C})  = \Bigg\lbrace
\begin{array}{c}
\,0,\,\,\,{\rm if} \,\, (z_{i_C},z_{j_C},z_{k_C}) \ \ {\rm satisfies \ \ clause\
\ C}\\ 
 \hskip.065in 1,\,\,\,{\rm if} \,\,(z_{i_C},z_{j_C},z_{k_C}) \ \ {\rm violates
\ \ clause\ \ C}\,.
\end{array}
\label{212}
\end{eqnarray}
We immediately turn these into quantum operators
\begin{equation}
H_\rP,_C(|z_1\rangle| z_2\rangle\cdots |z_n\rangle)=h_C(z_{i_C},z_{j_C},z_{k_C}
)|z_1\rangle | z_2\rangle\cdots |z_n\rangle
\label{213}
\end{equation}
and define
\begin{equation}
H_\rP =\sum_C H_\rP,_C.
\label{214}
\end{equation}
By construction $H_\rP$ is nonnegative, that is,
$\langle \psi|H_\rP|\psi\rangle\geq 0$ for all
$|\psi\rangle$ and $H_\rP|\psi\rangle =0$ if and only
if $|\psi\rangle$ is a superposition of states
of the form $|z_1\rangle
|z_2\rangle\cdots |z_n\rangle$ where the bit string 
$z_1 z_2  \ldots   z_n$ satisfies all of the clauses.  In
this context seeing if formula (\ref{21}) has a
satisfying assignment is accomplished by finding the
ground state of $H_\rP$.  If formula
(\ref{21}) has no satisfying assignment, the
ground state (or states) of $H_\rP$ correspond to the
assignment (or assignments) that violate the least
number of clauses.

$H_\rP$ given by Eq.~(\ref{214}) is the problem
Hamiltonian whose ground state we seek when we run
the quantum algorithm.  For the beginning
Hamiltonian $H_\rB$, first define
\begin{equation}
H_\rB^{(i)}=\half(1-\sigma_x^{(i)}).
\label{215}
\end{equation}
For each clause $C$, involving bits $i_C,j_C$, and
$k_C$, let 
\begin{equation}
H_\rB,_C =H_\rB^{(i_C)}+H_\rB^{(j_C)}+H_\rB^{(k_C)}\,.
\label{216}
\end{equation}
The beginning Hamiltonian is given by 
\begin{equation}
H_\rB=\sum_C H_\rB,_C.
\label{217}
\end{equation}
The ground state of $H_\rB$ is 
\begin{equation}
|x_1=0\rangle
|x_2=0\rangle\cdots |x_n=0\rangle=\frac{1}{2^{n/2}}\sum_{z_1}\sum_{z_2}\cdots 
\sum_{z_n}|z_1\rangle|z_2\rangle\cdots |z_n\rangle
\label{218}
\end{equation}
which we will take to be the initial state when we
run the quantum algorithm. 

The Hamiltonian that governs the evolution of the
quantum system is given by Eq.~(\ref{28}) with
$H_\rP$ specified in Eq.~(\ref{214}) and $H_\rB$
specified in Eq.~(\ref{217}).  Note that $H(t)$ is
a sum of terms, each of which is associated with
one of the clauses in (\ref{21}),
\begin{equation}
H(t)=\sum_C H_C(t)
\label{219}
\end{equation}
where
\begin{equation}
H_C(t)=(1-t/T)H_\rB,_C+(t/T)H_\rP, _C.
\label{220}
\end{equation}
Each $H_C(t)$ involves only the bits associated
with clause $C$ and therefore $H(t)$ is a sum of
terms each of which involves a fixed number of bits.

For a given problem such as EC3 or EC2 we must
specify the running time as a function of the
number of bits, $T(n)$.  Since the state
$|\psi(T)\rangle$ is not exactly the ground state
of $H_\rP$ we must also specify $R (n)$, the number
of times we repeat the quantum evolution in order
to have a desired probability of success.  This
paper can be viewed as an attempt to determine
$T(n)$ and $R(n)$ by numerical methods.

We now summarize the ingredients of the quantum
adiabatic evolution algorithm.  Given a problem
and an $n$-bit instance, we assume we know the
instance-independent running time $T(n)$ and
repetition number $R(n)$.  For the instance and
given $T(n)$,
\begin{enumerate}
 \item Construct the time-dependent Hamiltonian
$H(t)$ given by Eq.~(\ref{28}) with $H_\rB$ and
$H_\rP$ given by Eq.~(\ref{214}) and Eq.~(\ref{217}).
\item Start the quantum system in the state
$|\psi(0)\rangle$ given by Eq.~(\ref{218}).
\item Evolve according to Eq.~(\ref{24}) for a time
$T=T(n)$ to arrive at $|\psi(T)\rangle$.
\item  Measure $z_1,z_2,\ldots, z_n$ in the state
$|\psi(T)\rangle$ and check if  the bit string 
$z_1 z_2  \ldots   z_n$ satisfies all clauses.
\item Repeat $R(n)$ times.
\end{enumerate}

\setcounter{equation}{0}
\section{\large Choosing instances}
  \label{sec:3} 
For our numerical study we randomly  generate instances of
the two problems under study, EC3 and EC2.
Focus first on EC3. For a decision problem, it suffices
to produce a satisfying assignment when one or more exists. 
For now, we restrict our attention to instances with a
unique satisfying assignment.  We believe that
instances with only one satisfying assignment
include most of the difficult instances for
our algorithm.  In fact as we will see later, our
algorithm runs faster on instances with more than
one satisfying assignment so the restriction to a unique satisfying
assignment appears to restrict us to the most difficult cases.

With the number of bits fixed to be $n$, we
generate instances of EC3 as follows.  We pick
three bits at random, uniformly over the integers
from $1$ to $n$. (The bits must all be different.) 
We then have a formula with one exact cover
clause.  We calculate the number of satisfying
assignments.  We then add another clause by picking a new set of
three bits.  Again we calculate the number of
satisfying assignments. We continue adding clauses
until the number of satisfying assignments is
reduced to one or zero.  If there is one satisfying
assignment we accept the instance.  If there are
none we reject the instance and start again.  Using this procedure the
number of clauses is not a fixed function of the
number of bits but rather varies from instance to
instance.  For EC3 we find that  the number
of clauses is typically close to the number of bits.

We follow a similar procedure for EC2.  When we add
a clause we randomly specify which two bits are involved in
the clause. Again we repeat this procedure
until there are two satisfying assignments 
(or none in which case we discard the
instance).  We stop at two satisfying assignments because EC2 has a bit
negation symmetry. If $w_1w_2\ldots  w_n$ is a satisfying assignment
so is
$\bar{w}_1\bar{w}_2 \ldots \bar{w}_n$ and accordingly there are no instances with
a single satisfying assignment. For EC2 the number of clauses is typically close
to the number of bits.

We know that EC2 is classically computationally simple but of course there is
no guarantee that quantum adiabatic evolution will work well on EC2.  We
choose instances of EC2 with two satisfying assignments to parallel as
closely as possible our study of EC3.

\setcounter{equation}{0}
\section{\large Numerical simulation}
\label{sec:4}

We are exploring the quantum adiabatic evolution
algorithm by numerically simulating a perfectly
functioning quantum computer.  The quantum computer
takes an initial state $|\psi(0)\rangle$, given by
Eq.~(\ref{218}), and evolves it according to the
Schr\"{o}dinger equation (\ref{24}), for a time $T$,
to produce $|\psi(T)\rangle$.  The Hamiltonian
$H(t)$ is given by Eq.~(\ref{28}) with $H_\rP$ and
$H_\rB$ determined by the instances of satisfiability
being studied.

If the number of bits is $n$, the dimension of the
Hilbert space is $2^n$.  This exponential growth in
required space is the well-known impediment to
simulating a quantum computer with many bits.  For
our modest numerical investigation, using Macs
running MATLAB for a few hundred hours, we can only
explore out to 16 bits.  We integrate the
Schr\"{o}dinger equation using a variable step size
Runge-Kutta algorithm, checking accuracy by keeping
the norm of the state equal to unity to one part
in a thousand.  Since the number of bits is modest,
we can always explicitly determine the ground state (or ground states)
of $H_\rP$.  Given
$|\psi(T)\rangle$ and the ground state (or states) of $H_\rP$ we
can calculate the probability that a measurement of
$z_1,z_2,\ldots, z_n$ will give a satisfying
assignment by taking the sum of the squares of the inner
products of $|\psi (T)\rangle$ with $|w_1\rangle| w_2\rangle\cdots
|w_n\rangle$ where the bit strings $w_1w_2 \ldots  w_n$ are the satisfying
assignments.

\setcounter{equation}{0}
\section{\large The median time to achieve probability 1/8}
\label{sec:5}

Our goal in this paper is to explore the running time $T(n)$ and the repetition
number $R(n)$ that will give a successful algorithm.  To this end we first
determine the typical running time needed to achieve a fixed probability of
success for a randomly generated instance with $n$ bits for $7\leq n \leq 15$. 
In particular we determine the median time required to achieve a success
probability of $1/8$.  Since this is a numerical study we actually hunt for a
time that produces a probability between 0.12 and 0.13. For each $n$ between
7 and 15, for both EC3 and EC2, we find the median of 50 randomly
generated instances.

\numFig{1}{Exact Cover with three-bit clauses. Each circle is
the median time to achieve probability between 0.12 and 0.13 for
50 instances. The curve is a quadratic fit to the data.}{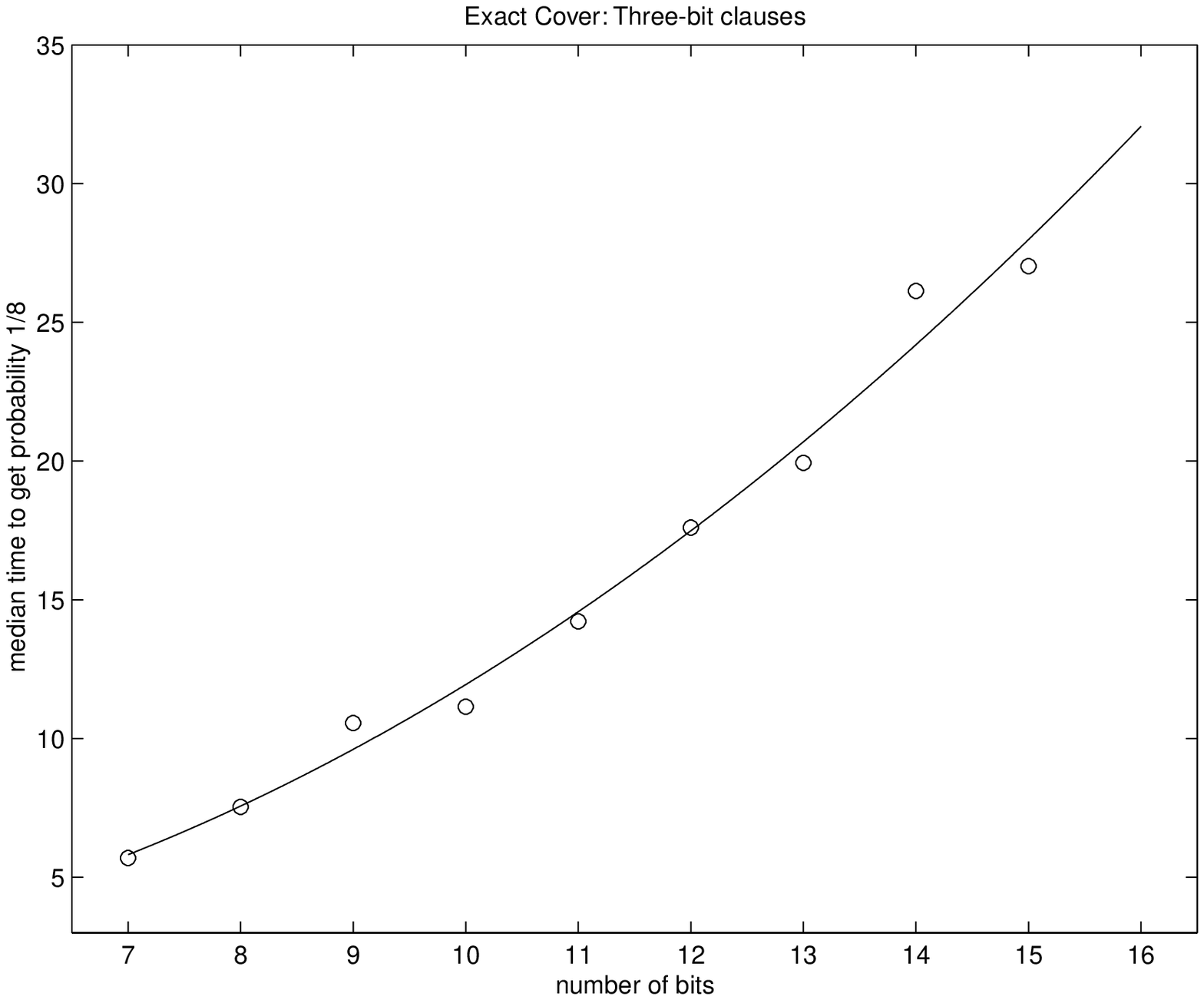 scaled 725}

In \refFig1 the circles represent the data for EC3 and the solid curve is a
quadratic fit to the data.  At each number of bits the times required to reach
probability $1/8$ range from roughly half the median to twice the median.  For
this range of $n$, a quadratic, or even linear fit, is clearly adequate.  The
exponential $1.618\,(1.215)^n$ is also a good fit. In the next section we show a
situation where an exponential fit to the data is
\emph{required} for the same range of $n$.

We know of one anomalous instance (discovered by Daniel Preda, outside of the
data collected for this paper) with 11~bits and a time to achieve probability 1/8
of close to 300. However, at $T=14.57$, which is the value of the quadratic fit in
\refFig1 at $n=11$, the probability of success for this instance is already 0.0606.
Because this probability is not anomalously low, the algorithm proposed in
Section~\ref{sec:7} will have no difficulty with this instance. 

\numFig{2}{Exact Cover with two-bit clauses. Each circle is
the median time to achieve probability between 0.12 and 0.13 for
50 instances. The curve is a linear fit to the data.}{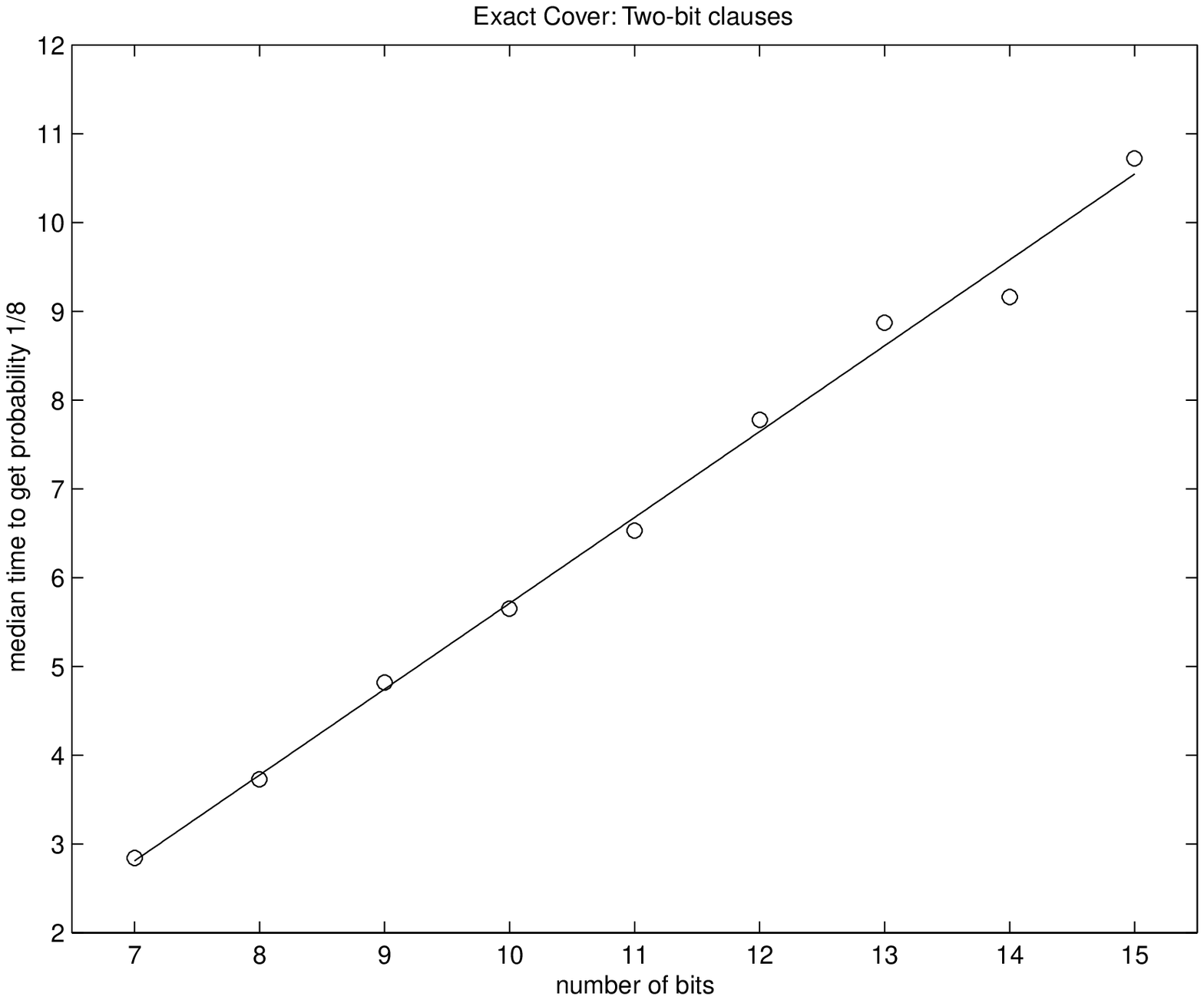 scaled 725}

In \refFig2 the circles represent the data for EC2 and the solid curve is a
linear fit to the data.  Here the maximum time required to reach probability
$1/8$, for each number of bits, is roughly $20\%$ greater than the
median.

\setcounter{equation}{0}
\section{\large  Destroying the bit structure of the Hamiltonian}
\label{sec:6}

The Hamiltonian $H(t)$ is a sum of terms $H_C(t)$ each of which involves only
the few bits mentioned in clause $C$; see Eq.~(\ref{219}). Each $H_C(t)$ is
associated with a subspace of the Hilbert space corresponding to the satisfying
assignments of clause $C$, that is, the space spanned by the ground states of
$H _\rP, _C$\,; see Eq.~(\ref{213}).  Quantum adiabatic evolution (for $T$ big
enough) yields a state in the intersection of the subspaces associated with all
of the clauses.  Our intuition is that the bit structure of these subspaces is
crucial to the success of the quantum algorithm.

To test this intuition we destroy the bit structure of the Hamiltonian and run
the algorithm again.  Specifically consider the classical energy function
that counts the number of violated clauses
\begin{equation}
h(z)=\sum_C h_C(z_{i_C},z_{j_C},z_{k_C})
\label{51}
\end{equation}
where $z=z_1z_2\ldots  z_n$ and $h_C$ is given in Eq.~({2.12}).  From
Eq.~(\ref{213}) and Eq.~(\ref {214}) we have that
\begin{equation}
H_\rP(|z_1\rangle|z_2\rangle\cdots  |z_n\rangle)=h(z)|z_1\rangle|z_2\rangle\cdots
|z_n\rangle.
\label{52}
\end{equation}
Now let
\begin{equation}
h_{\rS}(z)=h(\Pi(z))
\label{53}
\end{equation}
where $\Pi$ is a random permutation of the integers $\lbrace 0,1,2,\ldots,
2^n-1\rbrace$.  Note that $\Pi$ is {\it not} a permutation of the bits but
rather a random scrambling of all $2^n$ of the $z$'s.

Let
\begin{equation}
H_{\rS}, _\rP\,(|z_1\rangle
\,|z_2\rangle\cdots \,|z_n\rangle)\,=h_{\rS}\,(z)\, |z_1
\rangle\,|z_2\rangle\cdots  |z_n\rangle
\label{54}
\end{equation}
and accordingly
\begin{equation}
H_{\rS} (t) =(1-t/T) \,H_\rB \,+ \,(t/T) \,\, H_{\rS,\rP}.
\label{55}
\end{equation}
The spectrum of $H_{\rS,\rP}$ is identical to that of $H_\rP$ but the
relationship between the eigenvalues and the values of $z$ has been scrambled.
Accordingly the spectrum of
$H_{\rS}(t)$ is not equal to the spectrum of $H(t$) except at $t=0$ and $t=T$.  If
$H_\rP$ has a unique ground state so does $H_{\rS,\rP}$ and for $T$ large
enough we expect (again by the adiabatic theorem) that quantum evolution by
$H_{\rS}(t)$ will bring us to the ground state of $H_{\rS,\rP}$.

\numFig{3}{Exact Cover with two-bit clauses and a scrambled
problem Hamiltonian. Each circle is the median time to achieve probability
between 0.12 and 0.13 for 100 instances. The solid line on this log plot
represents an exponential fit to the data.}{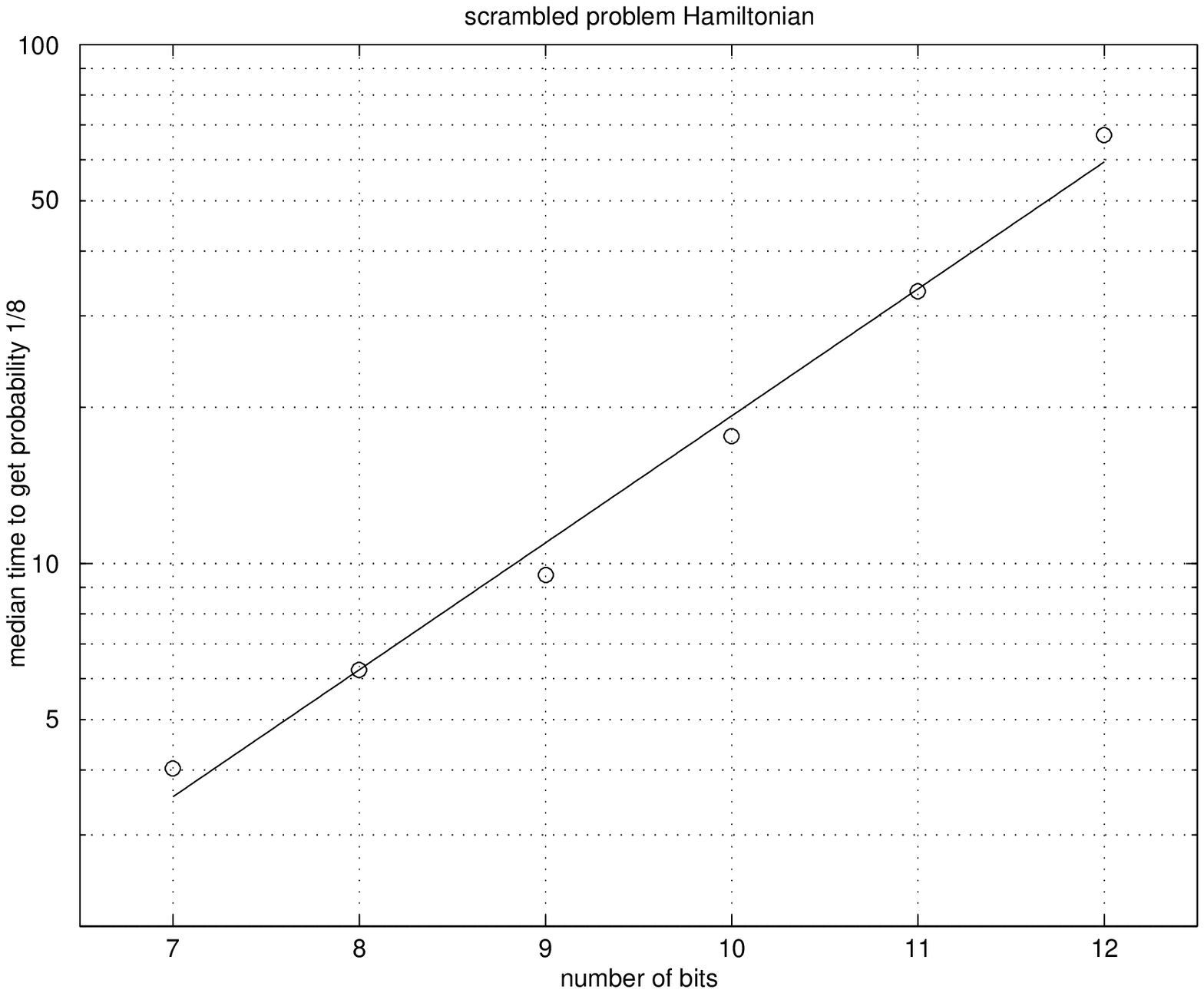 scaled 725}

Once the bit structure has been destroyed, finding the minimum of
$h_{\rS}(z)$ is essentially the problem of minimizing an arbitrary function
defined on $\lbrace 0,1,2,\ldots, 2^n-1\rbrace$.  Solving this problem requires
exponential time even on a quantum computer[2, 3].  To confirm this we generated
100 instances of EC2 for each of $n=7,8,9,10,11,12$.  For each instance we
generate a random permutation $\Pi$ and quantum evolve with $H_{\rS}(t)$ for
time $T$.  In \refFig3 we show the median time $T$ required to achieve a
success probability of $1/8$.  The linear behavior of the data on the log plot
indicates an exponential growth as a function of $n$ and the solid line
represents $0.0689\,(1.7565)^n$. 
In contrast with the data in Figures \ref{fig:1} and \ref{fig:2}, the data in
\refFig3 cannot be well fit by  a quadratic. 

\setcounter{equation}{0}
\section{\large  Probabilities of success at a proposed running time}
\label{sec:7}

Figures \ref{fig:1} and \ref{fig:2}, at least by comparison with \refFig3, indicate
that the median times required to achieve probability $1/8$, for EC3 and EC2
with
$7\leq n\leq 15$, grow modestly with $n$.  Thus the fitted medians, for EC3 and
EC2, are reasonable candidates for the running times $T(n)$ of our algorithm for
these two problems.  Automatically,  with these run times, our algorithm will
achieve a success probability of at least $1/8$ for about half of all instances.  Our
goal now is to explore how low the success probability can get at these run
times. To this end we generate 100 new instances of EC3 and EC2 for each $n$. 
Now
$n$ runs from 7 to 16 for EC3 and from 7 to 15 for  EC2  and $T(n)$ is
given by the fit to the data shown in Figures~\ref{fig:1} and~\ref{fig:2}.

\numFig{4}{Exact Cover with three-bit clauses. Probability of
success at the proposed running time. Circles represent the
medians of 100 instances at each number of bits. Triangles are the
10$^{\rm th}$ lowest probabilities and X's are the lowest.}{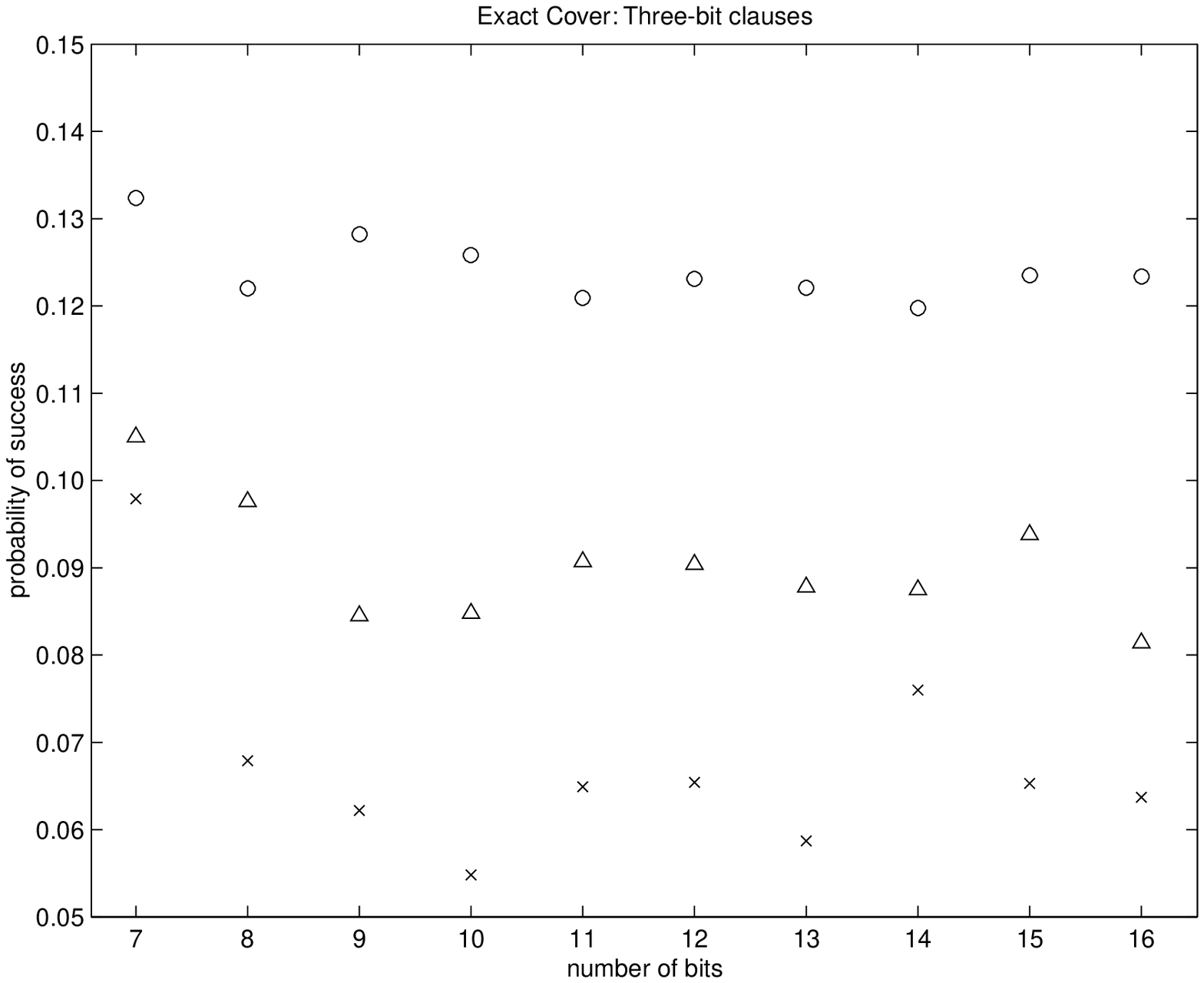 scaled 725}

\refFig4 displays the results for EC3.  For each value of $n$ from 7 to 16 we
show the median probability of success at $T(n)$ as well as the smallest of the
100 probabilities and the $10^{th}$ smallest.  It is no surprise that the
median probability is close to $1/8$ for all values of $n$.  The fact that the
smallest probability does not decrease with $n$ was not anticipated.  This
means that, at least for the range of the number of bits considered, the number
of repetitions $R(n)$ can be taken to be constant with $n$ to achieve a fixed
desired probability of success.  

\numFig{5}{Exact Cover with two-bit clauses. Probability of
success at the proposed running time. Circles represent the
medians of 100 instances at each number of bits. Triangles are the
10$^{\rm th}$ lowest probabilities and X's are the lowest.}{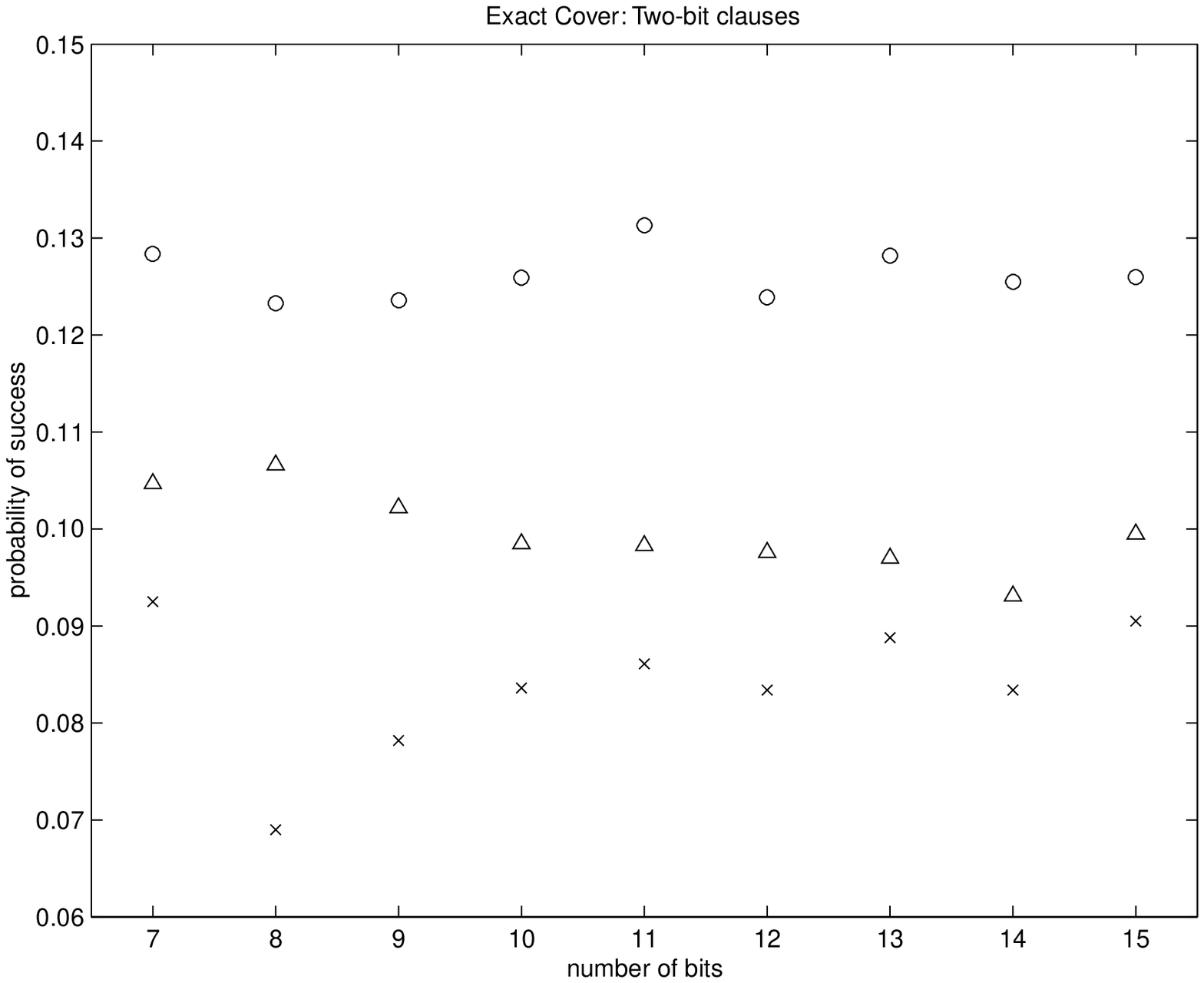 scaled 725}

\numFig{6}{Exact Cover with three-bit clauses running at $T=5.82$ for all
numbers of bits. Circles represent the median probability of 100
instances. The straight line behavior on the log plot shows an
exponential decrease in success probability.}{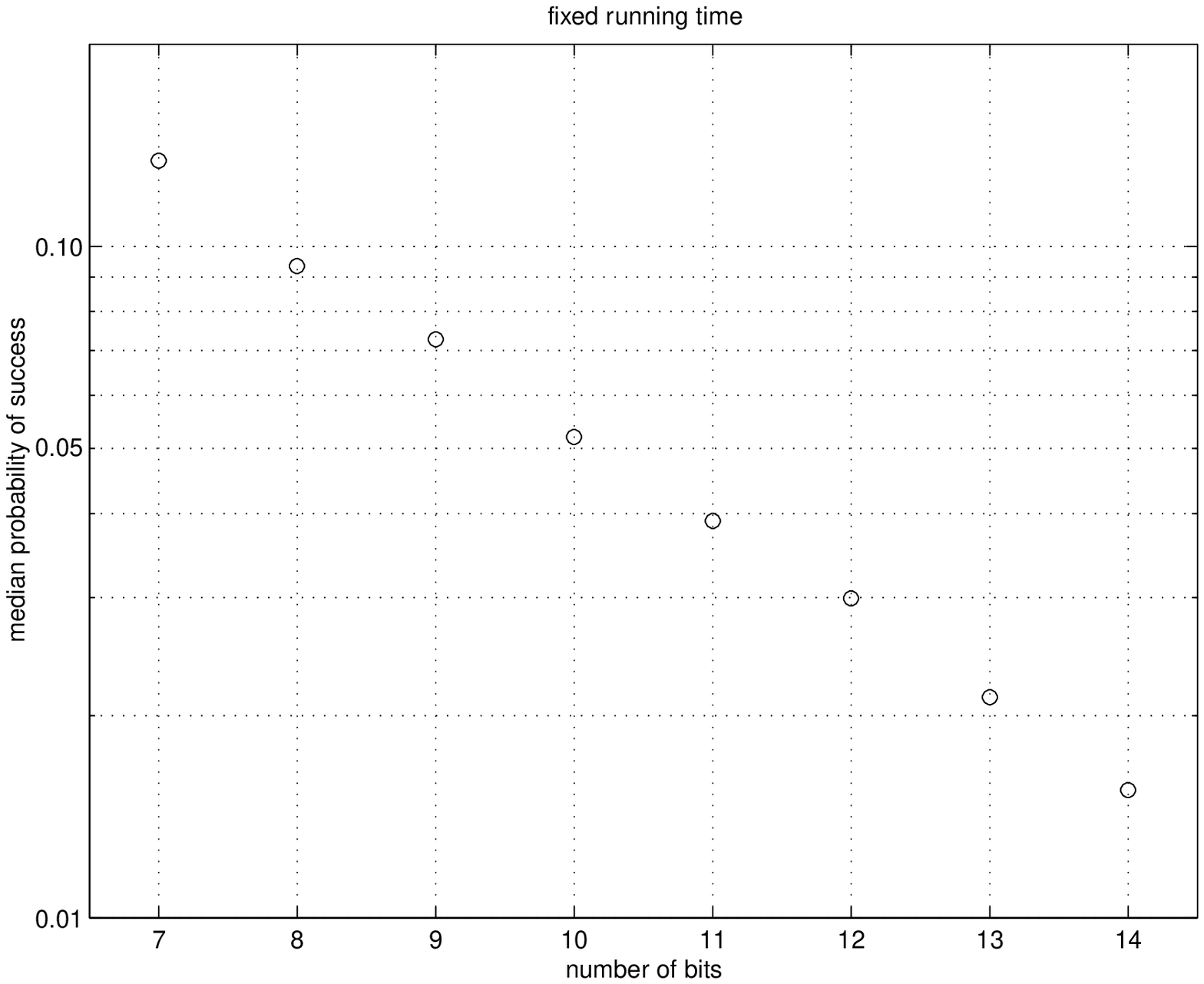 scaled 725}

In \refFig5 the data for EC2 is presented. Here the run time $T(n)$ is given
by the linear fit to the data in \refFig2.  Again the median probability at
each value of $n$ is close to $1/8$ as is expected.  However, even for this
classically easy problem we had no guarantee that the worst case probability
would not decrease with $n$. In fact it does not appear to decrease at all.

In order to show that the running time $T(n)$ must increase with $n$ to
produce a successful algorithm, we explore the success probabilities obtained
when using an $n$-independent running time.  More specifically for the EC3
instances used to generate \refFig4 we run the algorithm for $n=7,8,\ldots, 14$
for a constant value of $T$, the one previously used for $n=7$.  In \refFig6 the
log plot shows that the median success probability decreases exponentially,
which means that the number of repetitions $R(n)$ would have to grow
exponentially to achieve a fixed probability of success.

\numFig{7}{Exact Cover with three-bit clauses with a scrambled problem
Hamiltonian. The running time is given by the quadratic fit shown in \refFig1.
The  log plot shows an exponential decrease in success
probability.}{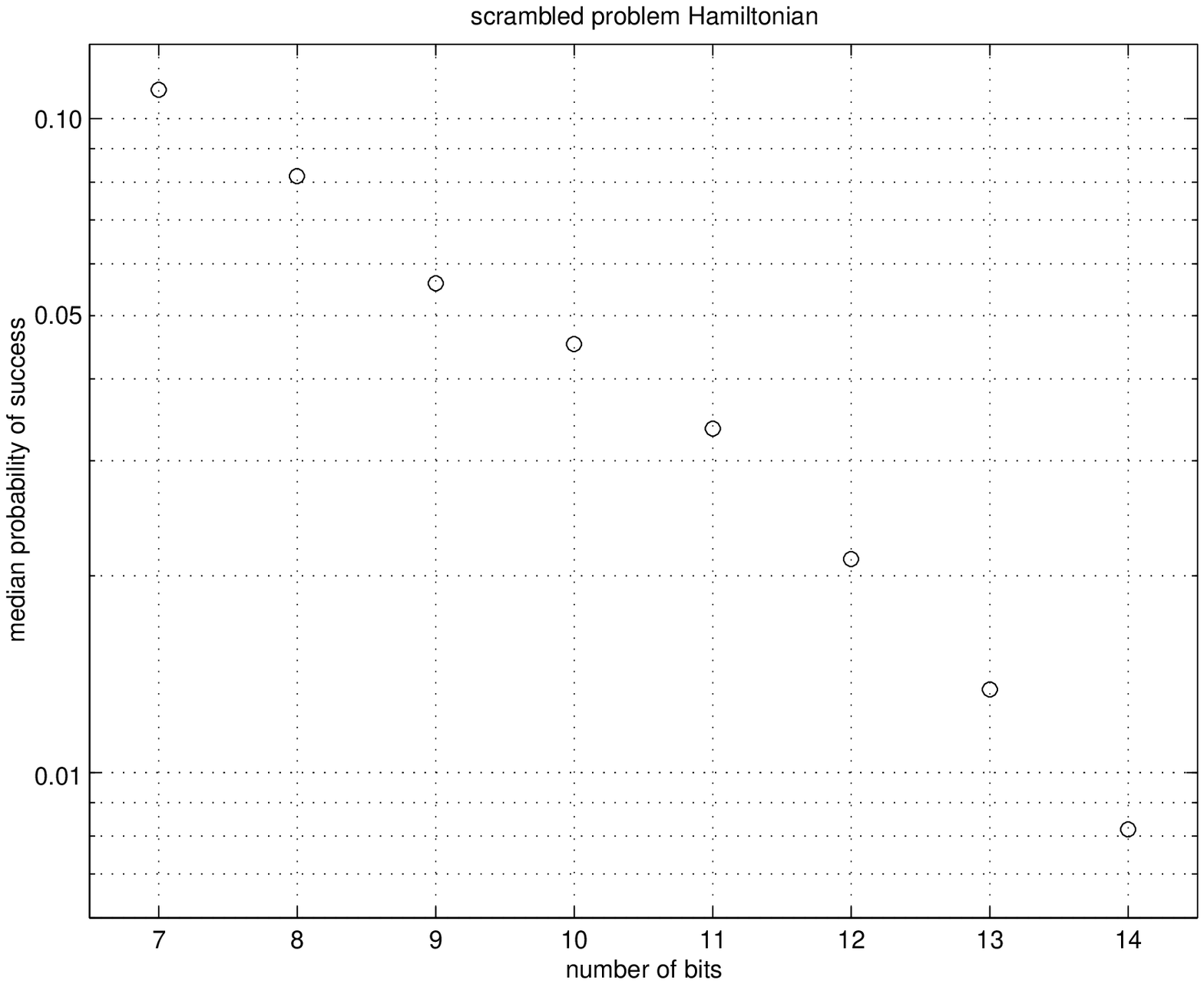 scaled 725}

In Section~\ref{sec:6} we gave evidence that the bit structure is crucial to the
success of the quantum adiabatic evolution algorithm.  We make this point
again by taking the instances of EC3, for $n=7,8,\ldots,14$, and running the
algorithm with
$T(n)$ given by the fit in \refFig1 but with $H(t)$ replaced by
$H_{\rS}(t)$.  In \refFig7, the median probability
of success is seen to decrease exponentially with $n$.  This helps confirm our
intuition that the quantum adiabatic evolution algorithm takes advantage of the
bit structure inherent in the problem.

\numFig{8}{Exact Cover with three-bit clauses and 6, 7, 8, or 9 satisfying
assignments. Circles represent the medians of 100 instances at each number of
bits. Triangles are the 10$^{\rm th}$ lowest probabilities and X's are the
lowest. Compare with \refFig4.}{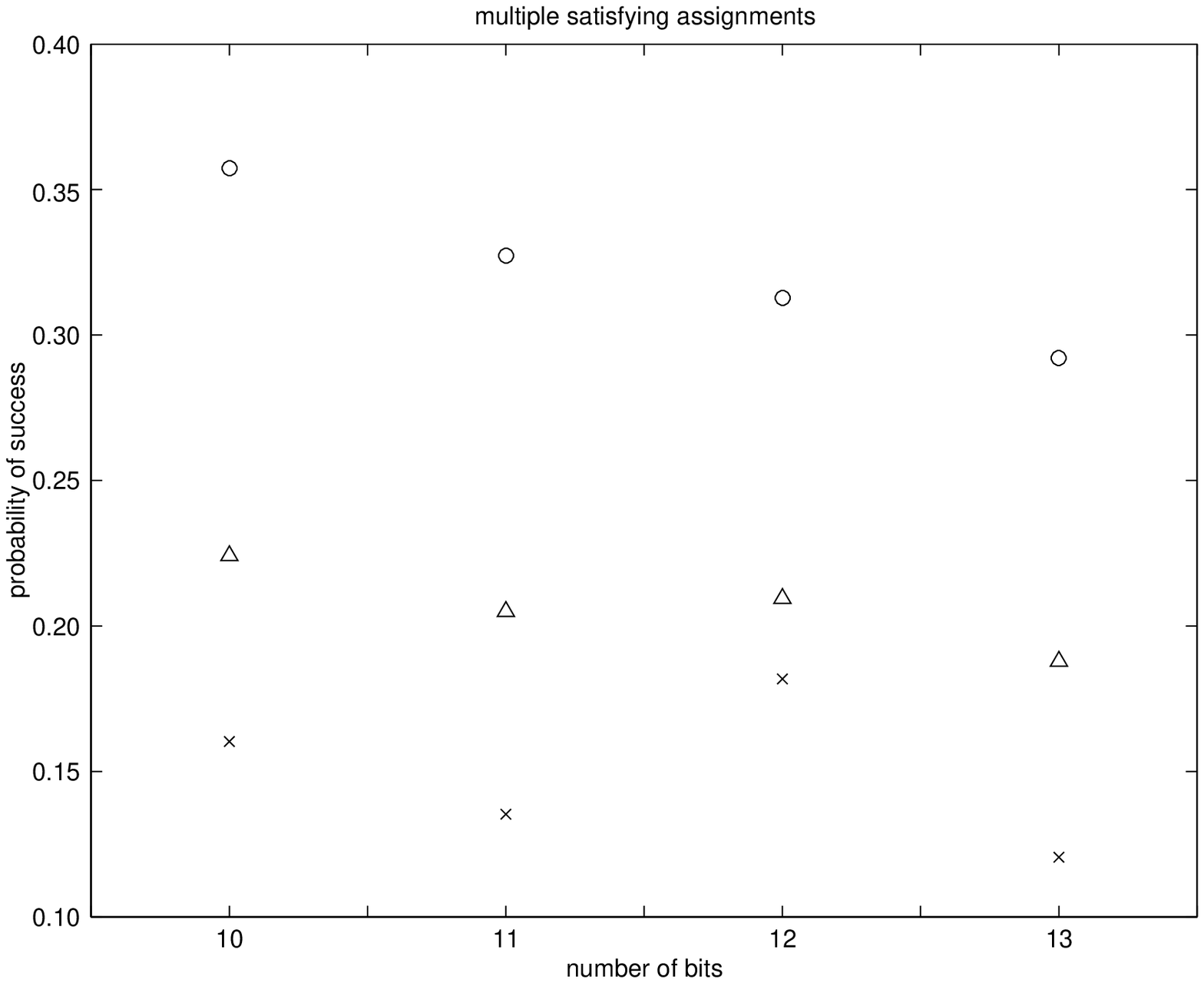 scaled 725}

\setcounter{equation}{0}
\section{\large  Instances of EC3 with more than one  satisfying
assignment}
\label{sec:8}

All of the EC3 data presented up to this point was generated from instances 
 with unique satisfying assignments.  Now we explore EC3 instances with
more than one satisfying assignment. As in Section~\ref{sec:3} clauses are
added at random but now instances are accepted as soon as the number of
satisfying assignments is 6, 7, 8, or~9.  If adding a clause reduces the
number of satisfying assignments from more than 9 to less than 6, the instance
is rejected.  We do this with 100 instances for 10, 11, 12, and 13 bits and run at
the same times $T(n)$ used for instances with a unique satisfying
assignment to generate \refFig4.  

In \refFig8 we show the median probability,
the smallest probability, and the $10^{\rm th}$ smallest for these instances. 
At the running times used, the median probability for instances with unique 
satisfying assignments is close to 1/8 (for any number of bits). For the instances
with multiple satisfying assignments the medians are about 1/3 and the worst
case  has a probability of about 1/8. This substantiates our intuition that
instances with unique satisfying assignments are generally the most difficult for
the quantum adiabatic algorithm.

At the running times explored in this paper transitions out of the instantaneous
ground state are not uncommon. In the case of a unique satisfying assignment
such a transition (assuming no transition back) leads to a final state that does not
correspond to the satisfying assignment. In the case of multiple satisfying
assignments such transitions may lead to states that are headed towards the
subspace spanned by the satisfying assignments. This is why the success
probabilities are typically higher when there is more than one satisfying
assignment.

\setcounter{equation}{0}
\section{\large Discussion}
\label{sec:10}

We have presented numerical evidence that a quantum adiabatic evolution
algorithm can solve instances of satisfiability in a time that grows slowly as a
function of the number of bits.  Here we have worked out to 16 bits, but with
more computing power instances with higher numbers of bits can be studied. 

This algorithm operates in continuous time.  The algorithm can be
written as a product of few-qubit unitary operators~\cite{ref:0} where the
number of factors in the product is of order $T(n)^2$ times a polynomial
in~$n$.   
 However,
understanding the idea behind the algorithm is obscure in the conventional
quantum computing paradigm.  (A quantum algorithm for satisfiability that is
explicitly within the ordinary paradigm is presented in~\cite{ref:3}.)  

The algorithm studied in
this paper works by having the quantum system stay close to the ground state
of the time-dependent Hamiltonian that governs the evolution of the system.  
We imagine that protecting a device that remains in its ground state from
decohering effects may be easier  than protecting a device that
requires the manipulation of excited states.

\subsection*{Acknowledgments}
This work was supported in
part by the Department of Energy under cooperative
agreement DE--FC02--94ER40818. E.F. thanks the participants at the Aspen
Center for Physics meeting on Quantum Information and  Computation (June
2000) for many helpful discussions. We thank Mehran Kardar, Joshua Lapan,
Seth Lloyd, Andrew Lundgren, and Daniel Preda for valuable input.

\end{document}